\documentclass[fleqn,usenatbib,useAMS]{mnras}
\usepackage{graphicx,slashbox}
\usepackage{amsmath}
\usepackage{amssymb}
\usepackage{bm}
\usepackage{color}
\usepackage{xcolor}

\title[PDF of a Molecular Clouds ensemble II]{Density distribution function of a self-gravitating isothermal compressible turbulent fluid in the context of Molecular Clouds ensembles II: the contribution of the turbulent term and the potential of the outer shells}

\author[Donkov, Stefanov]
{\parbox{\textwidth}{Sava Donkov$^{1}$, Ivan Stefanov$^{1}$}\vspace{0.4cm} \\
  $^1$Department of Applied Physics, Faculty of Applied Mathematics, Technical University, 8 Kliment Ohridski Blvd., 1000 Sofia, Bulgaria \\}

\begin{document}
\label{firstpage}

\date{Accepted XXX. Received YYY; in original form ZZZ}
\pagerange{\pageref{firstpage}--\pageref{lastpage}} \pubyear{2017}
\maketitle

\begin{abstract}
	
In this paper we continue to investigate the energy conservation equation obtained in our previous work. We set ourselves three new goals. The first one is to rewrite the main equations in terms of density profile in order to give more physical insight. The second one is to investigate the significance of two new terms in the energy conservation equation. They originate from the gravity of the outer shells of cloud and the masses outer to the cloud, respectively. The third goal is to investigate the main equation in the case when the kinetic turbulent term scales according to Larson's law and it is independent, formally, of the accretion, in contrast to the previous work. The combination of supersonic turbulence and spherical symmetry raise a caveat which is commented in our conclusions. We obtained two solutions for the density profile. They scale with slopes -2 and -3/2, respectively. The energy balance for the second solution is the same as in the previous paper: this is a free-fall. For the first solution there are two cases. The first one: if the turbulent term does not scale, then it could be important for the energy balance of the cloud. The second one: if the turbulent term does scale, then it is not important for the energy balance of the cloud. The two new gravitational terms don't affect the existence of the two solutions, but the gravitation of the outer masses calibrate the energy balance for the first solution.

\end{abstract}


\section{Introduction}   \label{Sec-Intr}

The understanding of the origin of probability distribution function (PDF) of mass density of the interstellar medium (ISM) is of great importance for the obtaining of an explanation of the star-formation process from first principles \citep{HF_12,Krh_14,KG_16}. The PDF of the medium is determined by the physics of the interstellar gas. On the other hand there is a link between the PDF and the local star-formation process. That is why the PDF of a given star-forming region is a tool of predicting of the initial stellar mass function, the star-formation rate, and the star-formation efficiency in this region \citep{Krh_14,Offner_ea_14}. If we know how the physics of a medium determines the PDF, then we can make a robust link between that physics and the local star-formation \citep{Elme_18}.

Our goal is to obtain the PDF from first principles. However, there are different physical regimes in the ISM. Recently \cite{DS_18} (hereafter Paper I) did an investigation of this task in the case of cold molecular gas with an isothermal equation of state \citep{Ferriere_01}. We studied a gas ball with radial symmetry which accretes material from the outside. The gas entering the cloud (our gas ball) through its boundary (with supersonic velocity) goes through all the scales and finishes in the centre of the ball where a very small and dense core is located (inside of which star-formation can eventually occur) (see \cite{Burkert_17}, so called simple bath-tube model). We also assumed that there is supersonic compressible turbulence and that it is locally homogeneous and isotropic in every shell of the gas ball. All the system is in steady state which concerns both: the macro-states (the motion of the fluid elements) and the micro-states (the thermal motion of the molecules). We neglected the magnetic fields and the back-reaction from new born stars. We also neglected the dissipation, assuming that our scales (these are the radii of the gas ball) belong to the inertial range of the turbulent cascade. So the physics of our system consists of gravity, supersonic turbulence and accretion, and thermodynamics (isothermal state). Solving the set of compressible Euler equations in spherical coordinates, after they were ensemble averaged, we obtained two equations. The first one, coming from the equation of motion of a fluid element, shows that the sum of the kinetic (accretion plus turbulent), the thermal and the gravitational energies of a fluid element per unit mass remains a constant when this fluid element moves through the scales (equation (9) in Paper I). The second one, coming from the continuity equation, gives a formula for the accretion velocity, expressed through the density and the scale (equation (15) there). Giving explicit forms for the energies per unit mass we solved the equations approximately up to the leading-order term in the series expansion, which assumes that the PDF is a power-law, in two cases: when the core is negligible (the fluid element is too far from the core) and when the core is important (the fluid element is near to the core). In the former case we obtain a solution with a slope of $-3/2$ (which counterparts to density profile with a slope of $-2$), this presumes a dynamical equilibrium, in the outer shells of the ball, between accretion and gravity. In the latter case we have a free-fall solution with a slope of $-2$ (which counterparts to density profile with a slope of $-3/2$) and a balance between the accretion and gravity of the core.

Our results correspond to previous studies. \cite{Larson_69} and \cite{Penston_69a} have investigated a collapsing homogeneous gas ball without accretion. The main forces are self-gravity and isothermal gas pressure. They have solved the equations of motion numerically and obtained a density profile with a slope of $-2$ in the outer layers. Also \cite{Shu_77} and \cite{Hunter_77} have treated the problem analytically and the first one has obtained two density profiles: $-2$ for the outer layers (but in static equilibrium: pressure supports against gravity), and $-3/2$ for the free-falling inner layers near to the singularity (the solution of \cite{Shu_77} describes the so called inside-out collapse). Using numerical simulations \cite{N-R_ea_15} have investigated a collapsing core embedded in a larger medium (called cloud) and accreting material from the latter. They have also obtained a density profile $-2$ in the outer layers of the core during its collapse in the cloud. Recently \cite{Li_18} has obtained density profile with a slope of $-2$ when gravity, accretion and turbulence interact. He claims that this slope is universal for scale-free gravitational collapse and that isothermal state is not a necessary condition. There are conformations also from observations. In systems like star cluster-forming molecular clumps, the radial density profile is very close to $-2$ \citep{Mueller_ea_02,Evans_03,Wyr_ea_12,Palau_ea_14,Wyr_ea_16,Csengeri_ea_17,Zhang-Li_17}.

In this paper, slightly changing the model, we set ourselves three main goals. The first one is to rewrite the main equations in terms of density profile. In Paper I we wrote the equations in the form to ask for probability density function as a unknown quantity, but we lost the physical insight. That is why we are going to eliminate this disadvantage. In addition we display the way in which our two solutions are obtained in clearer form. The second one is to reconsider our assumptions in Paper I, concerning the gravitational potential caused by the outer shells with respect to the position of the fluid element. In Section 3.2 (Paper I), where we discuss the explicit form of the gravitational term, we argue that only the gravitational potential which originates from the inner shells of the ball with respect to the position of the fluid element must be included in the equation. Our argument is that the outer shells do not contribute to the gravitational force. The latter is right, because the first derivative (taken with a negative sign), with respect to the radius, of the full gravitational potential is the force. The differentiation eliminates the contribution of the outer shells which means that the potential of the inner shells determines the motion of the fluid element through the scales. On the other hand we use the equation of balance of the energies per unit mass and we have to work with the full gravitational potential. In the present work we consider the influence of the outer shells of the cloud and conclude that the potential caused by them has negligible effect on the two solutions that we obtain. Adding a constant term, we also account for the gravitation of the masses outer to the cloud (this is a slight change of the model), assuming that they obey radial symmetry. In contrast to the unimportant outer shells of the cloud, the gravitation of the outer masses is significant for the energy budget in equations (\ref{EnEq-1.1}) and (\ref{EnEq-1.2}). The third goal is to investigate the main equation in the case when the kinetic turbulent therm is independent, formally, on the accretion and scales according to Larson's law. We suggested this equation in the discussion of Paper I (equation (28) there). This matters, because it is important to see if the two solutions do exist in the general case, not only in the particular one considered in Paper I.

The paper has the following structure. Section \ref{Sec-Equ_p(s)} is dedicated on derivation of the equation for density profile. In Section \ref{subsec-equ_rho(l)_terms} they are given the explicit forms of the terms in the above mentioned equation and there we account for the potential of the outer shells of the cloud with respect to the position of the fluid element in explicit form and also introduce the potential of the outer masses with respect to the cloud. After that, in Section \ref{subsec-equ_rho(l)_diff equ}, we obtain the equation for the density profile. We continue in Section \ref{Sec-analysis and sol} analysing the possible solutions of the latter equation and obtain them in two cases: far from the core (Section \ref{subsec-without G1}) and near to the core (Section \ref{subsec-with G1}). We discuss our results and give our conclusions in Section \ref{Sec-discussion}.

\section{Equation for the density profile}   \label{Sec-Equ_p(s)}

In this section we set ourselves the goal to rewrite the main equations in terms of the density profile $\varrho(\ell)$, an intrinsic characteristic of our cloud, and finally to obtain an equation which determines the latter quantity as an unknown function. In Paper I we derived the equation (20) which determines the quantity $Q(s)$, where $s=\ln(\rho/\rho_{\rm c})$ is the log-density, and $\rho_{\rm c}$ is the mass density at the outer boundary of the cloud. $Q(s)$ is simply the dimensionless cloud radius. In the present paper we denote the latter as $\ell$ and it takes values in the range $\ell_0\leq \ell \leq 1$, where the lower limit $\ell_0$ is the size of the small and dense core in the centre of our cloud, and the upper limit $1$ counterparts to the outer boundary of the cloud. For simplicity we use the dimensionless density profile $\varrho(\ell)= \rho(\ell)/\rho_{\rm c}$ which is a function of the dimensionless radius $\ell$, and is obviously the inverse function of $Q(s(\varrho))\equiv\ell(\varrho)$. Since we are also interested in obtaining of an expression for the PDF, the equation (\ref{p-rho}) gives the link between the latter and $\ell(\varrho)$.

Starting from the equations of the medium (see Section 3.1 in Paper I) under the assumption of steady state we obtain the equation for conservation of the total energy of a fluid element, per unit mass, during its motion through the cloud scales. This means the sum of the averaged kinetic, thermal and gravitational energies, per unit mass, is a constant with respect to $\ell$, or:

\begin{equation}
\label{dE/dl=0}
\frac{d}{d\ell}[\langle v^2/2 \rangle + \langle s \rangle + \langle \phi \rangle]=0~.
\end{equation}

\subsection{Explicit form of the terms in (\ref{dE/dl=0})}
\label{subsec-equ_rho(l)_terms}
In this subsection we derive the explicit form of the terms in equation (\ref{dE/dl=0}) taking into account the model presented in Paper I (Section 2) and also briefly reminded in Section \ref{Sec-Intr}. We start with the kinetic energy term:
\begin{equation}
\label{v=vt+va}
\langle v^2 \rangle = \langle v_{\rm t}^2 \rangle + \langle v_{\rm a}^2 \rangle~,
\end{equation}
where $\langle v_{\rm t}^2 \rangle$ is the turbulent kinetic energy per unit mass, and $\langle v_{\rm a}^2 \rangle$ is the accretion kinetic energy per unit mass. The proof that (\ref{v=vt+va}) is satisfied is given in Paper I (Section 3.2).

Our spherically symmetric cloud is ensemble averaged. That is why we choose to apply a standard scaling relation for  $\langle v_{\rm t}^2 \rangle$:
\begin{eqnarray}
\label{vt2-p(s)}
\langle v_{\rm t}^2 \rangle = \frac{u_0^2}{c_{\rm s}^2}\bigg(\frac{l_{\rm c}}{pc}\bigg)^{2\beta} \ell^{2\beta} = T_0 \ell^{2\beta}~,
\end{eqnarray}
where $u_0$ and $0\leq\beta\leq1$ are, respectively, the normalizing factor and the scaling exponent of the  turbulent velocity fluctuations in the standard law $u=u_0 L^{\beta}$ \citep{Larson_81, Pad_ea_06, Kritsuk_ea_07, Federrath_ea_10}. $T_0\equiv (u_0^2/c_{\rm s}^2)(l_{\rm c}/{\rm pc})^{2\beta}$ is the ratio of the turbulent kinetic energy per unit mass of the fluid element at the boundary of the cloud to the thermal energy per unit mass. This form of the turbulent kinetic energy per unit mass is different from the expression used in Paper I (equation 12 there), where the latter determines the dependence of the turbulence from the accretion. On the contrary, in this work, we presuppose that the turbulence is formally independent on the accretion. The explicit form of the accretion kinetic term has been obtained from the continuity equation in Paper I, Section 3.3, and it reads:

\begin{equation}
\label{va2-rho(l)}
\langle v_{\rm a}^2\rangle = A_0 \varrho(\ell)^{-2} \ell^{-4}~.
\end{equation}

 From the considerations in Paper I, Section 3.3 it stems that $\ell^{4}\varrho(\ell)^{2} \langle v_{\rm a}^2\rangle = const(\ell)=A_0$. Taking into account that the quantities are dimensionless and are normalized, respectfully, to the cloud size (for the scale), to the cloud edge density (for the density), and to the sound velocity (for the accretion velocity), we can obtain $A_0$ if we take $\ell=1$, that is at the cloud boundary. Then $\varrho=1$ and $\langle v_{\rm a}^2\rangle = u_{\rm a,c}^{2}/c_{\rm s}^{2}$. The latter is the ratio of the accretion kinetic energy term at the boundary of the cloud and the thermal kinetic energy per unit mass.

The thermal potential reads:  $\langle s \rangle = s$, since in our model the logarithmic density is averaged by assumption. The same is valid for the density: $\varrho=\langle\varrho\rangle$.

The averaged gravitational potential is given by the following expression:
\begin{equation}
\label{grav-potencial}
\langle \phi \rangle= -\frac{G}{l_{\rm c} c_{\rm s}^2}\frac{M(\ell)}{\ell} -\frac{G}{l_{\rm c} c_{\rm s}^2}\frac{M_0}{\ell} + \langle \phi^{\rm ext} \rangle~,
\end{equation}
 where $\ell$ is the radius at which the fluid element resides at the given moment, and $M(\ell)=3M_{\rm c}^{*}\int_{\ell_0}^{\ell} \ell'^2\varrho(\ell')d\ell'$ is the mass of the inner shells corresponding to $\ell$, where $M_{\rm c}^{*}= (4/3)\pi l_{\rm c}^3 \rho_{\rm c}$ is a normalizing coefficient the physical interpretation of which is given in Paper I (Section 3.2). Hence the first term in (\ref{grav-potencial}) is the gravitational potential caused by the shells which are inner with respect to the fluid element. $M_0$ is the mass of the dense core at the centre of the cloud and the second term in equation (\ref{grav-potencial}) is its gravitational potential at scale $\ell$. The last therm in (\ref{grav-potencial}) reads: $\langle \phi^{\rm ext} \rangle = -(3G M_{\rm c}^{*}/l_{\rm c} c_{\rm s}^2)\int_{\ell}^1 \ell'\varrho(\ell')d\ell' + \psi^{\rm ext}/c_{\rm s}^2$, where the first addend is the gravitational potential caused by the outer shells corresponding to $\ell$ and the second addend is the potential caused by the masses outside the cloud. For the latter we assume that all the masses outside the cloud give rise to potential $\psi^{\rm ext}$ in the volume of our cloud. And $\psi^{\rm ext}$ does not depend on the position of the fluid element during its motion through the scales. This, of course, is a simplification. Our assumptions are valid as long as the material outside the cloud obeys a radial symmetry. In reality this is not the case. This, however, is in agreement with the spirit of our model. Finally  $\langle \phi \rangle$ can be expressed by the density profile $\varrho(\ell)$:

\begin{eqnarray}
\label{grav-potencial_rho(l)}
\langle \phi \rangle= -\frac{3G}{c_{\rm s}^2}\frac{M_{\rm c}^{*}}{l_{\rm c}} \frac{\int\limits_{\ell_0}^{\ell} \ell'^2\varrho(\ell')d\ell'}{\ell} -\frac{3G}{c_{\rm s}^2}\frac{M_{\rm c}^{*}}{l_{\rm c}} \int\limits_{\ell}^1 \ell'\varrho(\ell')d\ell' \nonumber\\
 -\frac{G}{c_{\rm s}^2} \frac{M_0}{l_{\rm c}} \frac{1}{\ell}  + \psi^{\rm ext}/c_{\rm s}^2 ~.
\end{eqnarray}

\subsection{Derivation of the equation for $\rho(\ell)$}
\label{subsec-equ_rho(l)_diff equ}

With this preparation equation (\ref{dE/dl=0}) can be written in the following form:
\begin{eqnarray}
\label{dE/dl=0_rho}
\frac{d}{d\ell}\Bigg[A_0 \varrho(\ell)^{-2}\ell^{-4}+ T_0 \ell^{2\beta} + 2\ln(\varrho(\ell))\nonumber \\
- 3G_0 \frac{\int\limits_{\ell_0}^{\ell} \ell'^2\varrho(\ell')d\ell'}{\ell} -3G_0\int\limits_{\ell}^1 \ell'\varrho(\ell')d\ell' -\frac{G_1}{\ell}\Bigg]=0~,
\end{eqnarray}
where $G_0=(2G/c_{\rm s}^2)(M_{\rm c}^{*}/l_{\rm c})$ and $G_1=(2G/c_{\rm s}^2)(M_0/l_{\rm c})$ are dimensionless coefficients the physical meaning of which is clarified in Paper I, Section 3.4.

Let us denote the expression in the parentheses by $E_0$. This is the total energy per unit mass of the fluid element. It is clear that $\psi^{\rm ext}/c_{\rm s}^2$ contributes to $E_0$ and calibrates the total energy. If we compare the total energy of the fluid element in the present work ($E_0^{\rm II}$) and in Paper I ($E_0^{\rm I}$) we have the relation: $E_0^{\rm II} = E_0^{\rm I} - \psi^{\rm ext}/c_{\rm s}^2$ (for the role of the gravitational potential, caused by the outer masses, for the cloud's energy balance, see \cite{BP_ea_2018}).

Then we have:
\begin{eqnarray}
\label{E_rho}
A_0 \varrho(\ell)^{-2}\ell^{-4}+ T_0 \ell^{2\beta} + 2\ln(\varrho(\ell))\nonumber \\
- 3G_0 \frac{\int\limits_{\ell_0}^{\ell} \ell'^2\varrho(\ell')d\ell'}{\ell} -3G_0\int\limits_{\ell}^1 \ell'\varrho(\ell')d\ell' -\frac{G_1}{\ell}=E_0~.
\end{eqnarray}
Equation (\ref{E_rho}) is a non-linear integral equation for the function $\varrho(\ell)$. A  solution for $\varrho(\ell)$ would allow us to find the probability density function of mass density (if we know the inverse function $\ell(\varrho)$):
\begin{equation}
\label{p-rho}
{\rm PDF}(\varrho)=-3\ell(\varrho)^2 \frac{d\ell(\varrho)}{d\ln(\varrho)}~.
\end{equation}

\section{Study of the equation for the density profile}	
\label{Sec-analysis and sol}
We search for a solution of the form $\varrho(\ell)=\ell^{-p}$ which corresponds to a power-law PDF, $p(s)\propto \exp(qs)$ with $q=-3/p$ (see (\ref{p-rho})). The motivation for this ansatz is the same as in Paper I. A solution of this type is the simplest possible and besides the star-formation process occurs in the power-law tails of the PDFs.  A more general approach would be to ask for a solution in the form of a series of increasing exponents (with a small parameter $(1-\ell)$), but this is not our goal in the current work.

Making this substitution in (\ref{E_rho}) after some algebra we arrive at:
\begin{eqnarray}
\label{Ealg_rho}
A_0 \ell^{2p-4} + T_0 \ell^{2\beta} + 2(-p)\ln\ell \nonumber\\
 - 3G_0 \frac{\ell^{2-p}}{3-p} \Bigg[1-\bigg(\frac{\ell_0}{\ell}\bigg)^{3-p}\Bigg] - 3G_0 \frac{1-\ell^{2-p}}{2-p} - G_1 \ell^{-1} \nonumber\\
  = E_0~.
\end{eqnarray}
The expression on the left-hand side of the equation depends on $\ell$ which means that (\ref{Ealg_rho}) can be satisfied only approximately. Different assumptions and approximations yield different solutions for the parameter $p$ -  the slope of the  density profile.

We will study the following two cases. In the first case, the core can be neglected, i.e. $A_0,T_0, G_0 \gg G_1$, we search for a solution when the fluid element is far from the core ($1\gtrsim\ell\gg\ell_0$). In the second case, the core has a significant contribution or $A_0,T_0, G_0 \sim G_1$, the fluid element is near to the core ($\ell\sim\ell_0$).

\subsection{Solution far from the core}
\label{subsec-without G1}
When the core is neglected (\ref{Ealg_rho}) takes the form:
\begin{eqnarray}
\label{Ealg_rho_wotG1}
A_0 \ell^{2p-4} + T_0 \ell^{2\beta} + 2(-p)\ln\ell \nonumber\\
- 3G_0 \frac{\ell^{2-p}}{3-p} \Bigg[1-\bigg(\frac{\ell_0}{\ell}\bigg)^{3-p}\Bigg] - 3G_0 \frac{1-\ell^{2-p}}{2-p} = E_0 \ell^0 ~.
\end{eqnarray}
Before we continue the thermal term and the second addend in the parentheses of the gravitational therm resulted from the inner shells should be commented on. About the former: the turbulent and accretion velocities are supersonic by assumption, which means that the pressure term in the equation of motion (see equation (4), Paper I) is negligible in comparison to the kinetic terms. The thermal term in our equation comes from the pressure term, then it can be neglected. It can be important, possibly, only if the obtained solution for $p$ leads to exponents for accretion and turbulent therms which are positive. The second addend in the parentheses of the gravitational term is also negligible, because according to observations and simulations, typically,  $1\leq p \leq2$, and far from the core $\ell_0/\ell\ll1$, then $(\ell_0/\ell)^{3-p}\ll1$.

Then the exponents of the main terms, obtained with our ansatz, are respectively $2p-4,~2\beta,~2-p,~0$. An approximate solution of (\ref{Ealg_rho_wotG1}) can be obtained in the following way. With the approximations commented in the previous paragraph equation (\ref{Ealg_rho_wotG1}) contains only terms which have power law dependence on $\ell$. If the exponents of all the terms are equal then the powers of $\ell$ factor out and only constants  remain. The questions that arise then are which terms have  equal exponents and do they dominate over the rest of the terms? Since $0< \ell \leq 1$ the lower powers dominate over the higher ones. A non-trivial solution of (\ref{Ealg_rho_wotG1}) can be found only if the number of leading terms is at least two. If just one term dominates it remains unbalanced and the only solution is the trivial. In order to find a solution for $p$ we do the following. Choose a pair of terms and make the hypothesis that their exponents are equal and that the remaining terms are inferior or at most equal to them. Equating the two exponents we obtain a simple equation for $p$. We solve it, evaluate the exponents of all the terms with the obtained value and check if our hypothesis is confirmed. The same recipe is applied to all possible pairs of terms \citep{Zhivkov_99, RHB_2006}. 

Let us, for example, assume that the turbulent term and the accretion term have equal exponents and that they dominate over the other terms in equation (\ref{Ealg_rho_wotG1}). This assumption results in the following simple equation for $p$: $2\beta=2p-4$. Its root, $\beta+2$, is given in the first row and second column of table~1. 

As a next step, we use the obtained root for $p$ and evaluate the values of the exponents of all the terms in equation (\ref{Ealg_rho_wotG1}). The results are given in the second line of table~2. With this root just one of the terms, the one whose exponents is $-\beta$, dominates over the others. As it appears, the  assumption is not justified. Besides, the dominant term remains unbalanced. Hence, this root does not allow us to find a non-trivial solution and we will have to check the other possible pairs of exponents.

The roots for $p$ that we obtain with the above described procedure are:  $2,~\beta+2,~2(1-\beta)$. This is made clear in table 1. In table~2 the values of the exponents for every root are given. To make a conclusion about the existence of a solution of the equation (\ref{Ealg_rho_wotG1}) we have to remember the range of $\beta$: $0\leq\beta\leq1$. If $\beta=0$, the three cases are equivalent and there is only one solution: $p=2$ ($q=-3/2$), and the energy balance is: $$A_0+T_0-3G_0\approx E_0~.$$ If $\beta>0$, then there exist a solution only if $p=2$ ($q=-3/2$), and the energy balance is: $$A_0-3G_0\approx E_0~.$$

\begin{table}
	\centering
	\small
	\caption{Comparing each-other the exponents of the main therms in the equation (\ref{Ealg_rho_wotG1}), and the corresponding roots for $p$. }
	\begin{tabular}{|c|c|c|c|c|}
		\hline
		exponents & $2\beta$     & $2p-4$    & $2-p$       & 0\\
		\hline
		$2\beta$      & --            &  $\beta+2$ & $2(1-\beta)$& --\\
		\hline
		$2p-4$        & $\beta+2$    & --         & 2           & 2 \\
		\hline
		$2-p$         & $2(1-\beta)$ & 2         & --           & 2 \\
		\hline
	    0             & --            & 2         & 2           & -- \\
		\hline
	\end{tabular}\label{etiket}
\end{table}

\begin{table}
	\centering
	\small
	\caption{The values of the exponents of the main therms in the equation (\ref{Ealg_rho_wotG1}), according to every root obtained in table 1. }
	\begin{tabular}{|c|c|c|c|c|}
		\hline
	     \backslashbox{roots}{exponents}            	  & $2\beta$      & $2p-4$      & $2-p$       & 0\\
		\hline
		2                 & $2\beta$      & 0           & 0           & 0\\
		\hline
		$\beta+2$         & $2\beta$      & $2\beta$    & $-\beta$    & 0 \\
		\hline
		$2(1-\beta)$      & $2\beta$      & $-4\beta$   & $2\beta$    & 0 \\
		\hline
	\end{tabular}\label{etiket}
\end{table}

These approximate equalities express the balance of the energy components. They are valid only if and as long as the remaining terms can be neglected.

The last term on the left hand side in equation (\ref{Ealg_rho_wotG1}) deserves special attention. This term is the gravitational potential caused by the outer shells of the cloud with respect to the fluid element. If $p=2$ the denominator of this term equals to zero. But the numerator also vanishes. Applying the L'Hospital's Rule one can obtain a non-infinite limit: $$\frac{1-\ell^{2-p}}{2-p}\longrightarrow -\ln(\ell)~.$$ Then the entire gravitational term reads: $-3G_0(1-\ln(\ell))\simeq -3G_0$, because if the fluid element is far from the core, then $1\gtrsim\ell\gg\ell_0$ and $\ln(\ell)\sim 0$.

Moreover, when $p=2$, if one takes into account the considerations in Paper I, Section 4.1, concerning the average density of the hole cloud, then $3G_0=\langle G \rangle$. This is the averaged gravitational energy per unit mass of the fluid element for the entire cloud. Similar property have the terms for the accretion kinetic and turbulent kinetic energies: $A_0=\langle A \rangle$ and $T_0=\langle T \rangle$, because accretion does not scale if $p=2$ and the turbulent term is important only if it does not scale, too.

Finally, when the core is neglected there exists only one solution: $\varrho(\ell)=\ell^{-2}$  (the PDF is: ${\rm PDF}(s)\approx (3/2) \exp(-3s/2)$), but there are two possibilities for energy balance. The first one, if the turbulence does not scale ($\beta=0$), reads:

\begin{equation}
\label{EnEq-1.1}
\langle A \rangle + \langle T \rangle - \langle G \rangle \approx E_0~.
\end{equation}

 And the second one, if the turbulence scales ($\beta>0$), is:

 \begin{equation}
 \label{EnEq-1.2}
 \langle A \rangle - \langle G \rangle \approx E_0~.
 \end{equation}

\subsection{Solution near to the core}
\label{subsec-with G1}
When the core is not negligible (the fluid element is near to the core) equation (\ref{Ealg_rho}) reads:
\begin{eqnarray}
\label{Ealg_rho_wtG1}
A_0 \ell^{2p-4} + T_0 \ell^{2\beta} + 2(-p)\ln\ell \nonumber\\
- 3G_0 \frac{\ell^{2-p}}{3-p} \Bigg[1-\bigg(\frac{\ell_0}{\ell}\bigg)^{3-p}\Bigg] - 3G_0 \frac{1-\ell^{2-p}}{2-p} - G_1 \ell^{-1} \nonumber\\
 = E_0\ell^0~.
\end{eqnarray}

According to the same arguments like in the previous section we can neglect the thermal term. The gravitational term, accounting for the potential of the inner shells, is also $\sim0$, because near to the core $\ell\sim\ell_0$, and the expression in the parentheses vanishes. Hence the exponents of the main terms are: $2p-4,~2\beta,~2-p,~-1,~0$. We can apply the same method for obtaining the solutions for $p$ like in the previous section, but there is a simpler physical consideration. If the core is important, then the leading order exponent must be $-1$. In this case there are two possibilities. The first one is the gravitation of the core to be balanced by the gravitational term resulting from the outer shells of the cloud and $2-p=-1$ leads to $p=3$. But the energy balance fails, because both terms are negative. The second one, and the only possible, is the gravitation of the core to be balanced by the accretion term, and it requires $p=3/2$. Therefore the only solution in that case is $\varrho(\ell)=\ell^{-3/2}$ (the PDF is: ${\rm PDF}(s)\approx 2 \exp(-2s)$) and the energy balance is:

\begin{equation}
\label{EnEq-2}
A_0 - G_1 \approx 0 ~.
\end{equation}

This is the well-known from Paper I free-fall solution (see Section 4.2 there).

\section{Discussion and conclusions}
\label{Sec-discussion}

In the previous sections we wrote the main equations in the terms of density profile and obtained more physical insight expressions. The latter gives the model more clarity. In addition, using table 1 and table 2, we illustrated the method of obtaining the solutions of the equation (\ref{Ealg_rho}), which is clarified also in the text of Section \ref{subsec-without G1}. With this we consider the first goal as accomplished.

The second one was to investigate the equation (\ref{dE/dl=0}) in the case when we account for the gravitational potential caused by the outer shells with respect to the position of the fluid element, in contrast to the previous work. According to the considerations in Section \ref{subsec-without G1} and Section \ref{subsec-with G1} one can conclude that the gravitation of the outer shells of the cloud is not important for the two cases that we have studied. So our two solutions are not influenced by the new term on the left hand side in the equation (\ref{E_rho}). This is not the case with the constant gravitational therm $\psi^{\rm ext}/c_{\rm s}^2$ caused by the masses outer to the cloud. It contributes to the total energy of the fluid element $E_0$ and we believe that it will be of key importance for the right energy balance in equations (\ref{EnEq-1.1}) and (\ref{EnEq-1.2}) (\cite{BP_ea_2018}).

Thinking on the third problem: to present the kinetic turbulent energy in a more general form, we have to note that the both solutions we have obtained in this work are the same as the solutions in Paper I. But there are differences in the equations of energy balance. For the second solution $\varrho=\ell^{-3/2}$ the energy balance has preserved its form as in Paper I. This is a free-fall and the energy balance per unit mass for the fluid element reads: $ A_0 - G_1 \approx 0 $.

But for the first solution $\varrho=\ell^{-2}$ there are two regimes. If $\beta=0$, then the turbulent kinetic energy could be important: $ \langle A \rangle + \langle T \rangle - \langle G \rangle \approx E_0 $ (this holds only if $T_0\sim A_0$). The lack of scaling reminds us of coherent cores \citep{Goodman_ea_1998}, whose scales are of order $\ell\sim 0.1~{\rm pc}$. This phenomenon ($\beta=0$) is observed, also, at larger scales in Rosette molecular cloud (see \cite{Vel_ea_18}, Section 5.4). If $\beta> 0$, then the turbulent kinetic energy is not important: $ \langle A \rangle - \langle G \rangle \approx E_0 $. The latter case strongly supports the idea for hierarchical and chaotic gravitational collapse at all of the cloud scales \citep{BP_ea_2011a,BP_ea_2011b,I-M_ea_2016,BP_ea_2018,Elme_18}.

We shortly remind that our model is an attempt of an abstract statistical description of classes of molecular clouds whose PDF, size $\ell_{\rm c}$, core size $\ell_0$, edge density $\rho_{\rm c}$, core density $\rho_0$, and temperature $T$ are the same. The ball (which is the average representative of the class), obeying radial symmetry, is an idealized object, but this is the simplest one that we could construct. It is clear that we have lost the specific morphology and physics of every cloud from the class, but we believe we catch the main properties of the class members.

The compatibility of supersonic turbulence and spherical symmetry is the major caveat of our model. Large scale supersonic turbulence gives rise to shocks which might result in substantial departures from the spherical symmetry. Such departures will induce a non-symmetric gravitational potential and hence will have influence over gravitational therms in our equation. It is difficult to say precisely how these departures will affect the energy balance equations, but we consider our approach as a first step in this task.

Another significant problem is that the second solution, near to the core, is a free-fall with a profile $p=3/2$ ($q=-2$). Some modern simulations \citep{KNW_11} and observations \citep{Schneider_ea_15}, where two power-law tails occur, report a different profile for the second tail: $p\sim 3$ ($q\sim -1$). This value can be explained, according to authors of the cited papers, by a decrease of the mass flow rate (fall under the action of gravity) from the larger to the smaller scales of the cloud. Among the possible reasons for such a decrease are: non-zero angular momentum of the small dense structures (of the dense core in the centre of the cloud, in our case) \citep{KNW_11,Schneider_ea_15}, large opacity and, respectively, pressure increase as a consequence of temperature increase (i.e. the system leaves the isothermal regime), the presence of magnetic fields, the back-reaction on the cloud from the newborn stars \citep{Schneider_ea_15} etc. All this physics is neglected in our model, that is why the inconsistency between the second slopes seems normal. This is a hint of the possible directions of elaboration of the model. One way is to suggest that near to the core the system leaves the isothermal regime and we have a polytropic equation of state: $p_{\rm th}\propto \rho^\Gamma$ and $\Gamma\neq 1$. This change in thermodynamics leads to the following equation near to the core:
\begin{eqnarray}
\label{Ealg_rho_wtG1_polytropic}
A_0 \ell^{2p-4} + T_0 \ell^{2\beta} + \frac{\Gamma}{\Gamma-1} \ell^{p(1-\Gamma)} \nonumber\\
- 3G_0 \frac{\ell^{2-p}}{3-p} \Bigg[1-\bigg(\frac{\ell_0}{\ell}\bigg)^{3-p}\Bigg] - 3G_0 \frac{1-\ell^{2-p}}{2-p} - G_1 \ell^{-1} \nonumber\\  = E_0\ell^0~.
\end{eqnarray}

We leave the study of this equation as a work for the future.

\section*{Acknowledgements}

This research is partially supported by the Bulgarian National Science Fund under Grant N 12/11 from 20 December 2017.

\bibliographystyle{mnras}

\end{document}